# Hardware implementation of photonic neuromorphic autonomous navigation

Yonghang Chen[1], Shuiying Xiang[1*], Xintao Zeng[1], Mengting Yu[1], Tao Zou[1],

Shangxuan Shi[1], Xingxing Guo[1], Yanan Han[1], Yahui Zhang[1*], and Yue Hao[1]

[1]*State Key Laboratory of Integrated Service Networks, Xidian University, Xi'an 710071, China.*

**ABSTRACT: Reinforcement learning (RL) is a core technology enabling the transition of artificial intelligence (AI) from perception to decision-making, but its deployment on conventional electronic hardware suffers from high latency and energy consumption imposed by the von Neumann architecture. Here, we propose a photonic spiking twin delayed deep deterministic policy gradient (TD3) reinforcement learning architecture for neuromorphic autonomous navigation and experimentally validate it on a distributed feedback laser with a saturable absorber (DFB-SA) array. The hybrid architecture integrates a photonic spiking Actor network with dual continuous-valued Critic networks, where the final nonlinear spiking activation layer of the Actor is deployed on the DFB-SA laser**

array. In autonomous navigation tasks, the system achieves an average reward of 58.22±17.29 and a success rate of 80%±8.3%. Hardware-software co-inference demonstrates an estimated energy consumption of 0.78 nJ/inf and an ultra-low latency of 191.20 ps/inf, with co-inference error rates of 0.051% and 0.059% in task scenarios with and without obstacle interference, respectively. Simulations for error-activated channels show full agreement with the expected responses, validating the dynamic characteristics of the DFB-SA laser. The architecture shows strong potential for integration with large-scale photonic linear computing chips, enabling fully-functional photonic computation and low-power, low-latency neuromorphic autonomous navigation.

## 1. Introduction

Reinforcement learning (RL) is a class of machine learning methods in which agents learn optimal policies through continuous interaction with their environment. Its theoretical foundation combines the trial-and-error learning mechanism from behavioral psychology with the reward optimization principle from optimal control theory [1-4]. As a core paradigm for sequential decision-making, RL has driven major advances across domains such as autonomous driving [5-7], strategic gaming [8-11], robotic control [12-14], and many other applications [15-20]. Recent algorithmic developments, including Twin Delayed Deep Deterministic Policy Gradient (TD3) [21], Proximal Policy Optimization (PPO) [22], Deep Deterministic Policy Gradient (DDPG) [23], and Deep Q-Network (DQN) [15], have significantly improved the capability of RL to handle high-dimensional state spaces, enhance sample efficiency, and stabilize policy learning. However, the deployment of these algorithms on conventional von Neumann architectures remains limited by the intrinsic separation of memory and computation, which introduces substantial latency and energy overheads, critically constraining the real-time performance of RL in applications such as autonomous navigation [24].As an emerging paradigm for high-performance computing, photonic computing has demonstrated remarkable potential due to its ultrafast signal propagation and inherent parallelism [25-28]. By utilizing photons instead of electrons as information carriers, photonic computing enables highly energy-efficient and low-latency linear operations through the intrinsic speed and parallel nature of light [29-43]. Existing photonic linear computing implementations primarily include microring resonator (MRR)-based weight banks [31-33], cascaded Mach-Zehnder modulators (MZMs) [34, 35], phase-change material (PCM) crossbar arrays [36, 37], and Mach-Zehnder interferometer (MZI) meshes [38-43]. However, despite recent advances [44-46], realizing a fully-functional photonic computing system remains challenging—particularly in implementing nonlinear operations. Against this backdrop, photonic spiking neural networks (PSNNs) have emerged as a promising computational framework that integrates the high-speed, energy-efficient processing of photonic computing with the event-driven and sparse dynamics of SNNs [47-55]. In 2019, Feldmann et al. experimentally demonstrated the first fully integrated optical spiking neuromorphic network based on PCMs, achieving pattern recognition on optically encoded letters with an energy consumption of approximately 60 pJ per synaptic weight update [47]. In 2020, Robertson et al. reported a photonic spiking neuron based on a vertical-cavity surface-emitting laser (VCSEL), achieving ultrafast detection of both oriented and omnidirectional edges in digital images [48]. In 2021, Xiang et al. proposed an all-optical SNN entirely based on VCSELs for supervised learning, theoretically demonstrating optical digit classification while eliminating the laterally inhibitory mechanism that is difficult to realize in optical systems [49]. In 2022, Lee et al. proposed and implemented an event-driven optoelectronic spiking neuron based on the Izhikevich model, achieving 97% accuracy and an energy efficiency of 50 trillion operations per second per joule (TOPS/J) on the MNIST handwritten digit recognition task [50]. In 2023, Xiang et al. fabricated and experimentally validated a photonic spiking neuron chip based on a Fabry-Perot laser with a saturable absorber (FP-SA), which realized single-neuron pattern classification through hardware-software co-computation [51]. In the same year, the team further developed and demonstrated a photonic synaptic core chip based on a DFB-SA laser, which, for the first time, simultaneously implemented linear weighting and nonlinear spiking activation within a single device, achieving 87% accuracy on the MNIST dataset [52]. More recently, Talukder et al. constructed a large-scale PSNN comprising 40,000 neurons, achieving bio-inspired excitability using a modified Ikeda map with slow inhibitory feedback. The network reached 77.5% accuracy on MNIST using only 8.5% of the neurons and represented the first hardware-compatible demonstration of simultaneous perturbation stochastic approximation (SPSA) training [53]. These advances collectively demonstrate the superior energy efficiency, processing speed, and sparsity of PSNNs. Nevertheless, their application to autonomous navigation remains unexplored. Gao et al. introduced a photonic spiking neuron-based obstacle-avoidance system employing a FP laser, which used rate coding to emulate biological monocular vision and experimentally verified the feasibility of ultrafast, low-power robotic obstacle avoidance [54]. However, the decision-making process of this system relies on predefined thresholds, lacking autonomous interaction and online learning between the RL agent and the environment, thus preventing full-process neuromorphic autonomous navigation. Integrating the ultrafast perception-decision capability of PSNNs with the sequential decision-optimization ability of RL is therefore expected to provide a promising pathway toward intelligent, self-adaptive, low-power and low-latency neuromorphic autonomous navigation systems.

This paper presents a photonic spiking TD3 reinforcement learning architecture for neuromorphic autonomous navigation, enabling highly energy-efficient intelligent decision-making through hardware-software co-inference. The main contributions of this work are as follows:

1. We propose a photonic spiking TD3 reinforcement learning framework that integrates a spiking-based Actor network with dual continuous-valued Critic networks. The algorithm is optimized under the hardware constraints of the photonic neuron chips, where the Actor network adopts positive-weight and bias-free structures to ensure hardware compatibility. In autonomous navigation tasks, the proposed architecture achieves an average reward of 58.22±17.29 and a navigation success rate of 80%±8.3% in software simulation, demonstrating the feasibility of employing PSNNs for autonomous navigation.

2.   The final nonlinear spiking activation layer of the Actor network is deployed on a DFB-SA laser array. The system achieves an estimated inference energy of 0.78 nJ/inf and an ultra-low latency of 191.20 ps/inf, highlighting the intrinsic low-power and low-latency advantages of photonic computing for reinforcement learning.

3.   Hardware-software co-inference experiments are conducted under two scenarios—with and without obstacle interference. The co-inference error rates are 0.051% and 0.059%, respectively, validating the high accuracy and reliability of PSNN-based reinforcement learning for neuromorphic autonomous navigation. Additionally, simulations based on the Yamada model using algorithm-generated inputs for error-activated channels show full agreement with the expected spiking responses, thereby theoretically validating the dynamic characteristics of the DFB-SA laser.

## 2. Methods

### 2.1. Photonic Spiking TD3 Reinforcement Learning Architecture

The proposed photonic spiking reinforcement learning architecture is based on the TD3 algorithm, an enhanced variant of the DDPG method. TD3 is designed to mitigate the overestimation bias of Q-values introduced by function approximation errors in the Actor-Critic framework, thereby enhancing both algorithmic stability and overall performance [21]. This is achieved through three key innovations: employing twin Critic networks (clipped double Q-learning), introducing target policy smoothing regularization, and adopting delayed policy updates. These mechanisms collectively stabilize training, improve policy optimization accuracy, and enhance performance in complex continuous control tasks. Owing to its superior stability and scalability, TD3 has been widely applied in domains requiring precise decision-making within high-dimensional state-action spaces, including robotic control [56, 57], autonomous driving [58, 59], and Unmanned Aerial Vehicles (UAV) communications [60, 61].

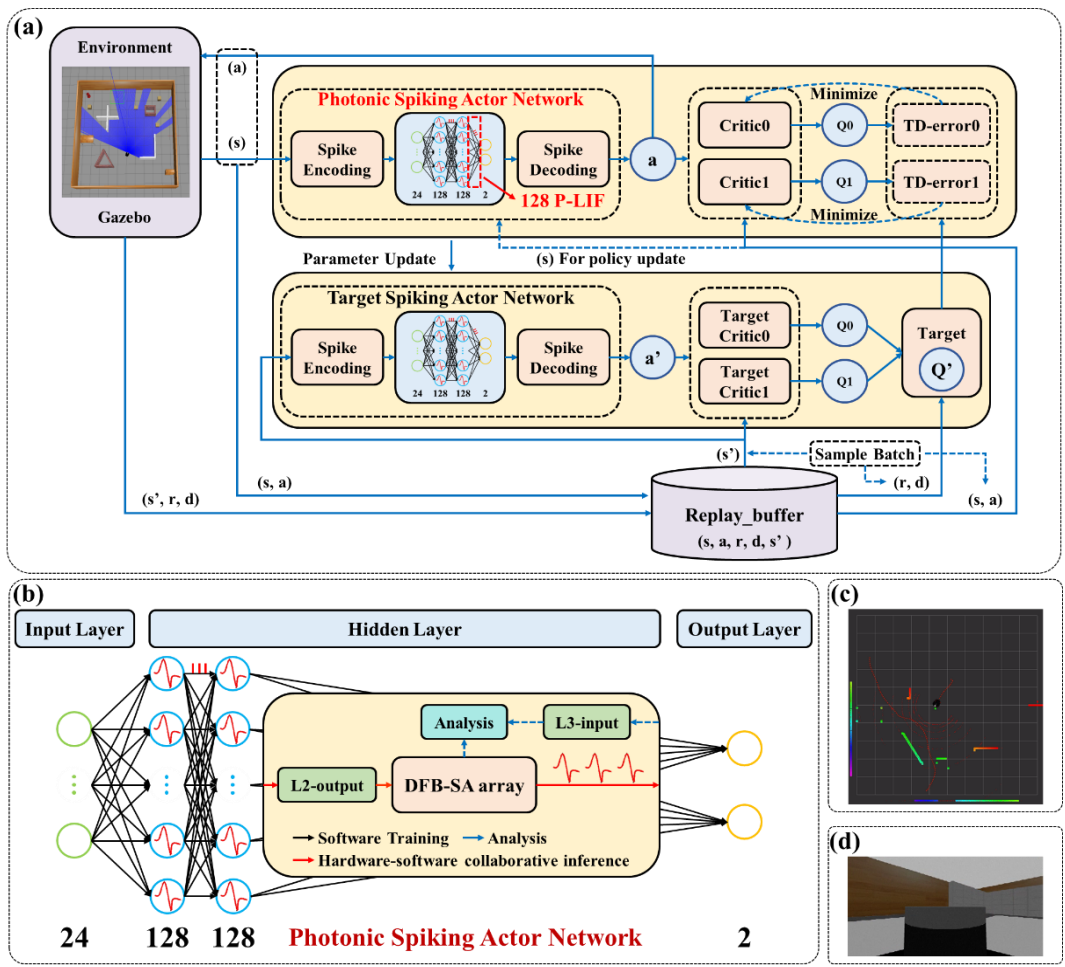

**Figure 1**. Overall architecture and implementation details of the photonic spiking TD3 reinforcement learning system. (a) Overall architecture of the photonic spiking TD3 reinforcement learning system. (b) Hardware-software co-inference framework. (c) Real-time navigation path map. (d) Onboard camera view of the robot.

Figure 1(a) illustrates the proposed photonic spiking TD3 reinforcement learning architecture. The reinforcement learning agent comprises a PSNN-based Actor network and dual Artificial Neural Network (ANN)-based Critic networks. The photonic spiking Actor network generates two continuous actions—the linear and angular velocities of the robot—based on the current state, while the Critic network evaluates the corresponding state-action Q-values. Using an ANN-based Critic offers more stable gradient estimation and faster convergence, which complements the event-driven and energy-efficient characteristics of the photonic spiking Actor network, resulting in a balanced learning framework. We considered a continuous control task in the Gazebo simulation environment [62], where the robot perceives obstacles through Light Detection and Ranging (LiDAR) sensors and performs mapless dynamic obstacle avoidance and goal-directed navigation through reinforcement learning. The overall system comprises three core modules: the agent, the Gazebo simulation environment, and the differential-drive control system, interconnected through the Robot Operating System (ROS) framework to enable LiDAR-based, mapless dynamic navigation [63]. Detailed configurations of the simulation environment, state space, and action space are provided in SI-1.

In the autonomous navigation tasks, the photonic spiking Actor network processes real-time state data from the Gazebo simulation environment. The static state inputs are first transformed via spike encoding by expanding them along the temporal dimension and replicating them over T timesteps to match the time-series processing paradigm of SNNs. The encoded sequences are then propagated through the photonic spiking Actor network, where each hidden layer employs leaky integrate-and-fire (LIF) neurons to produce temporal spiking responses. These spiking outputs are subsequently decoded by temporal aggregation (i.e., averaging over time) to recover static representations, which are normalized using a Tanh function to generate continuous velocity commands. During training, Gaussian noise is injected into the actions to encourage exploration. In addition, a forced exploration strategy is adopted: when LiDAR detects nearby obstacles, the agent executes random actions with a certain probability to escape local optima. After each interaction, the environment returns rewards, next states, and termination signals, which are stored in a replay buffer. For policy evaluation, mini-batches randomly sampled from the replay buffer are processed by dual-ANN Critic networks (800×600×1) with a twin-Q architecture. Target networks are employed to stabilize Q-value estimation by minimizing the temporal-difference error. Target actions are generated by an independent target photonic spiking Actor network with clipped noise to mitigate Q-value overestimation; this target Actor follows the same spike encoding–LIF activation–spike decoding pipeline as the primary Actor. The Actor network is updated every two training steps by maximizing the expected state–action value using gradients provided by the Critic. This delayed policy update strategy, together with soft target network synchronization, ensures stable and efficient convergence. The pseudocode of the complete algorithm workflow is provided in SI-2.

## 2.2. Hardware-Aware Algorithm Optimization and Photonic Implementation

To ensure hardware compatibility and general applicability, the photonic spiking

Actor network is co-designed with hardware-aware constraints for deployment on generic photonic linear computing platforms. While current photonic linear processors have experimentally demonstrated matrix operation scales up to 128×128 [43], we configure the Actor network to 24×128×128×2 as a representative and hardware-compatible example. This configuration illustrates the scalability and implementability of our approach within the capabilities of existing photonic linear computing architectures, while preserving high control accuracy. During software pretraining, the hidden-layer weights are constrained to be positive, and all bias terms are removed, ensuring that the trained parameters can be directly mapped onto photonic linear computing hardware. The final nonlinear spiking activation layer is deployed in hardware using a DFB-SA laser array, which performs optical nonlinear computation through laser spiking dynamics. This design enables seamless integration with large-scale photonic matrix-vector multiplication chips, realizing a fully-functional photonic computing architecture in which both linear and nonlinear operations are executed entirely in the optical domain. Simulation results verify the performance and stability of this architecture, providing theoretical guidance for subsequent hardware deployment and co-inference experiments.

## 2.3. Hardware-Software Co-Inference Architecture

To validate the effectiveness of the proposed photonic spiking TD3 reinforcement learning architecture, a hardware-software co-inference scheme is employed for the photonic spiking Actor network. Figure 1(b) presents the overall co-inference architecture. The process begins with software pretraining, where the photonic spiking Actor network is trained from scratch using a surrogate gradient method with a single time step (T = 1) to obtain a pretrained model that already incorporates the hardware constraints outlined in Section 2.2. Following pretraining, the final nonlinear spiking activation layer is deployed in hardware using a DFB-SA laser array, which realizes optical nonlinear computation via laser spiking dynamics. The corresponding input-output data from the pretrained software model are fed into the photonic hardware to evaluate its response. The experimentally measured optical outputs are compared with the software-generated counterparts to quantify the co-inference error rate, validating the consistency between optical and digital inference. Figures 1(c) and 1(d) show the real-time navigation trajectory and onboard camera view of the robot, respectively, providing an intuitive visualization of the simulation environment and the autonomous navigation behavior of the robot.

## 2.4 Experimental Setup

The final nonlinear spiking activation layer of the photonic spiking Actor network is experimentally deployed using a DFB-SA laser array. The fabrication detail of the DFB-SA laser is like that reported in Ref. [64]. Figure 2(a) illustrates the experimental setup of this layer, including a micrograph of the DFB-SA laser chip used in the experiment. In the schematic, red lines denote optical paths, while blue dashed lines correspond to electronic paths.

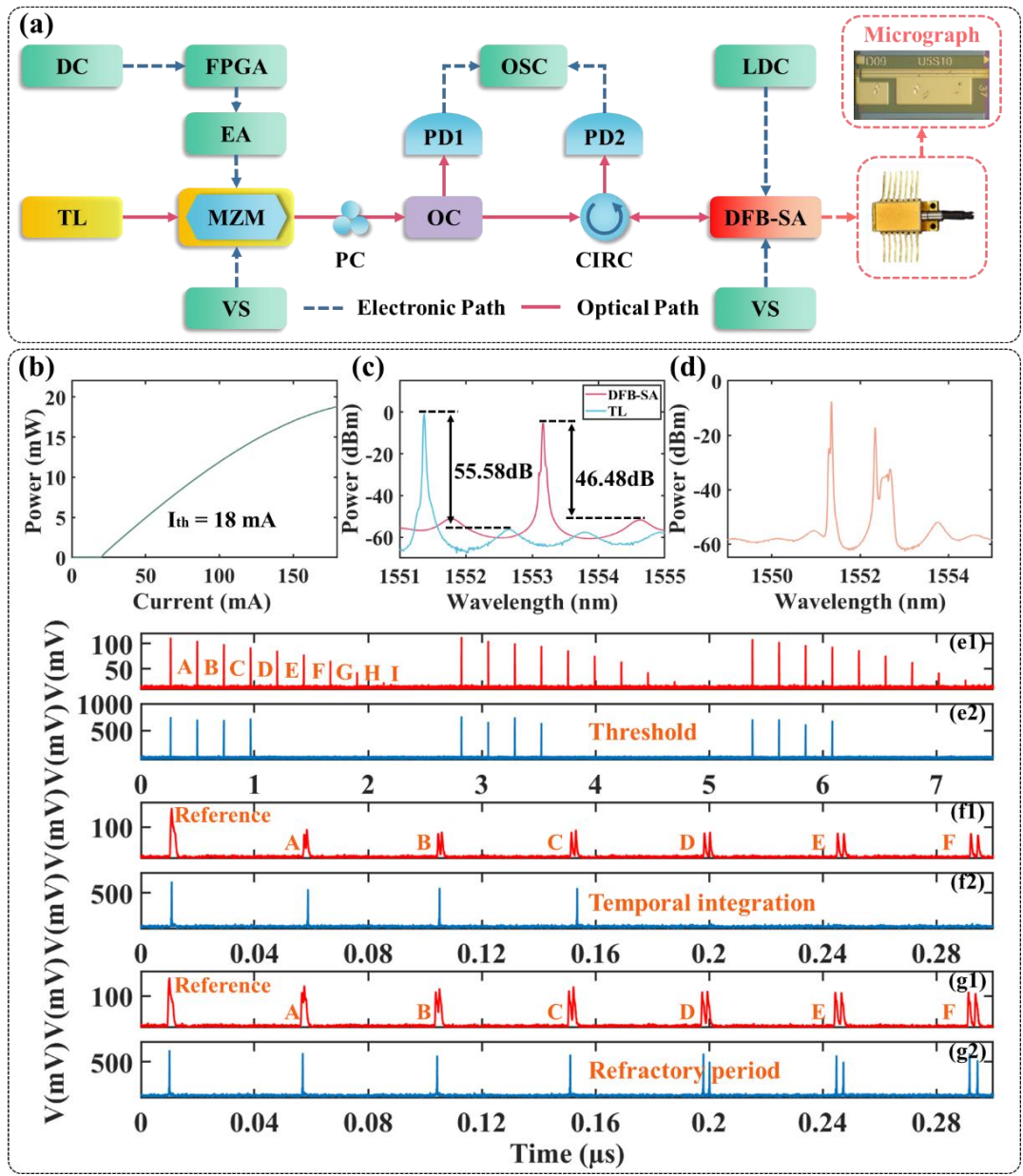

**Figure 2.** Experimental setup and properties of a DFB-SA laser-based photonic spiking neuron. (a) Experimental setup of the final activation layer in the photonic spiking Actor network. DC: digital computer; FPGA: field-programmable gate array; EA: electronic amplifier; TL: tunable laser; MZM: Mach-Zehnder modulator; VS: voltage source; PC: polarization controller; OC: optical coupler; PD: photodetectors; OSC: oscilloscope; CIRC: circulator; LDC: laser diode controller. (b) The PI curve of DFB-SA laser. (c) The optical spectra of the TL and DFB-SA laser. (d) The optical spectrum of the DFB-SA laser. (e1) - (g2) The neuron-like response of the DFB-SA laser.

A tunable laser (TL) provides the optical carrier, which is injected into the Mach-Zehnder modulator (MZM, Fujitsu FTM7928FB), while a voltage source (VS) supplies the driving voltage. Electronic signals generated by a field-programmable gate array (FPGA), controlled by a digital computer (DC), are amplified through an electronic amplifier (EA) and injected into the MZM for modulation. The modulated optical signal is then injected into the DFB-SA laser via a three-port optical circulator (CIRC), with a polarization controller (PC) adjusting the polarization state. A laser diode controller (LDC, ILX Lightwave LDC3724B) provides high-precision temperature control and low-noise bias currents to the gain region of the DFB-SA laser, while a voltage source applies a reverse bias voltage to the saturable absorber. The output of the DFB-SA laser is detected by a photodetector (PD, Agilent HP11982A) and recorded with an oscilloscope (OSC, Keysight DSOZ592A) and is subsequently analyzed along with the modulated optical signal tapped from the optical coupler (OC). Figure 2(b) shows the PI curve of the DFB-SA laser, with a threshold current of 18 mA. Figure 2(c) shows the spectra of the free-running tunable laser and the DFB-SA laser, from which the side-mode suppression ratios (SMSRs) are measured to be approximately 55.58 dB and 46.48 dB, respectively, demonstrating excellent single-mode performance and beam quality, which are critical for the experiment. Figure 2(d) shows the output spectrum of the DFB-SA laser with injections generated from TL. Figures 2(e1)-(g2) demonstrate the key neuro-mimetic dynamics of the DFB-SA laser—threshold, temporal integration, and refractory period—thereby validating its capability to emulate complex photonic spiking neuron behaviors. During the experiment, the DFB-SA laser temperature is maintained at 22 °C, the reverse bias voltage of the saturable absorber is set to 0.76 V, the bias current of the gain region is set to 27.9 mA, and the injection power $P_{inj}$ is 97.96 µW.

## 2.5 Theoretical Model of DFB-SA Laser Based on Yamada Model

In 2002, Minoru Yamada proposed a theoretical analysis of self-sustained pulsation phenomena and investigated the application of rate equations in narrowband-structured semiconductor lasers. The model can be used to analyze the dynamic characteristics of the DFB-SA laser [65]. Here, we modify the Yamada model to simulate a photonic spiking neuron based on the DFB-SA laser. The mathematical formulation of the Yamada model can be expressed as follows:

$$\frac{dN_1}{dt} = \frac{I}{eV_1} - \frac{N_1}{\tau_s} - \frac{\Gamma_1 g_1}{V_1}(N_1 - N_{g_1})S \tag{1}$$

$$\frac{dN_2}{dt} = -\frac{N_2}{\tau_s} - \frac{\Gamma_2 g_2}{V_2}(N_2 - N_{g_2})S \tag{2}$$

$$\frac{dS}{dt} = [\Gamma_1 g_1(N_1 - N_{g_1}) + \Gamma_2 g_2(N_2 - N_{g_2})]S - \frac{S - S_{ext}}{\tau_{ph}} + \beta_{sp}\frac{N_1 V_1}{\tau_s} \tag{3}$$

Here, subscripts 1 and 2 refer to the gain region and saturable absorber, respectively. $N$ denotes the carrier density, $g$ the gain coefficient, $\Gamma$ the confinement factor, $V$ the cavity volume, and $N_g$ the transparence carrier density. $S_{ext}$ denotes the number of externally injected photons, $\beta_{sp}$ is the spontaneous emission coefficient, $\tau_{ph}$ is the photon lifetime, $\tau_s$ is the carrier lifetime, and $I$ represents the bias current applied to the gain region. For simplicity, carrier diffusion between the gain region and saturable absorber is not considered in the model. The corresponding parameter values are listed in Table 1.

**Table 1.** The corresponding parameter values of DFB-SA laser

| Parameters | Gain region | Saturable absorber |
|---|---|---|
| Gain coefficient  $g$ | $1.8794 \times 10^{11}$ m$^3$/s | $0.594 \times 10^{-11}$ m$^3$/s |
| Confinement factor  $\Gamma$ | 0.08 | 0.08 |
| Cavity volume  $V$ | $172.8 \times 10^{-18}$ m$^3$ | $7.2 \times 10^{-18}$ m$^3$ |
| Transparence carrier density  $N_g$ | $1.4 \times 10^{24}$ m$^{-3}$ | $1.6 \times 10^{24}$ m$^{-3}$ |
| Photon lifetime  $\tau_{ph}$ | $2.56 \times 10^{-3}$ ns | $2.56 \times 10^{-3}$ ns |
| Carrier lifetime  $\tau_s$ | 1 ns | 1 ns |

## 3. Results

In this section, we present the experimental results of neuromorphic autonomous navigation using the photonic spiking TD3 reinforcement learning architecture with hardware-software co-inference. Figures 3(a1) and 3(a2) show the average reward curve and the navigation success rate curve of the baseline software algorithm, respectively. Since the training curves exhibit oscillations after convergence, the mean and standard deviation are computed over the last 20% of the training data, yielding an average reward of 58.22±17.29 and a success rate of 80%±8.3%. The environment configuration and training parameters used in the algorithm are provided in SI-3. After convergence, two task scenarios—with and without obstacle interference—are selected for experimental implementation of the final nonlinear spiking activation layer in the photonic spiking Actor network. Because the initial and target positions of the robot are randomized, the number of state-action sample pairs varies between tasks. Demonstration videos for both task scenarios (with and without obstacle interference) are provided in Supplementary Video 1 and Supplementary Video 2, respectively.

Figures 3(b1) and 3(b2) illustrate the start and goal positions, complete navigation trajectories, and the distributions of discrepancies between the expected and experimentally measured spiking activation outputs, where a value of 1 represents a false-positive spike (overactivation) and -1 represents a false-negative spike (missed activation). The experimental results confirm that the activation outputs of the DFB-SA array closely match the expected activations. For clarity, the scenario without obstacle interference is referred to as Task 1 and that with obstacle interference as Task 2.

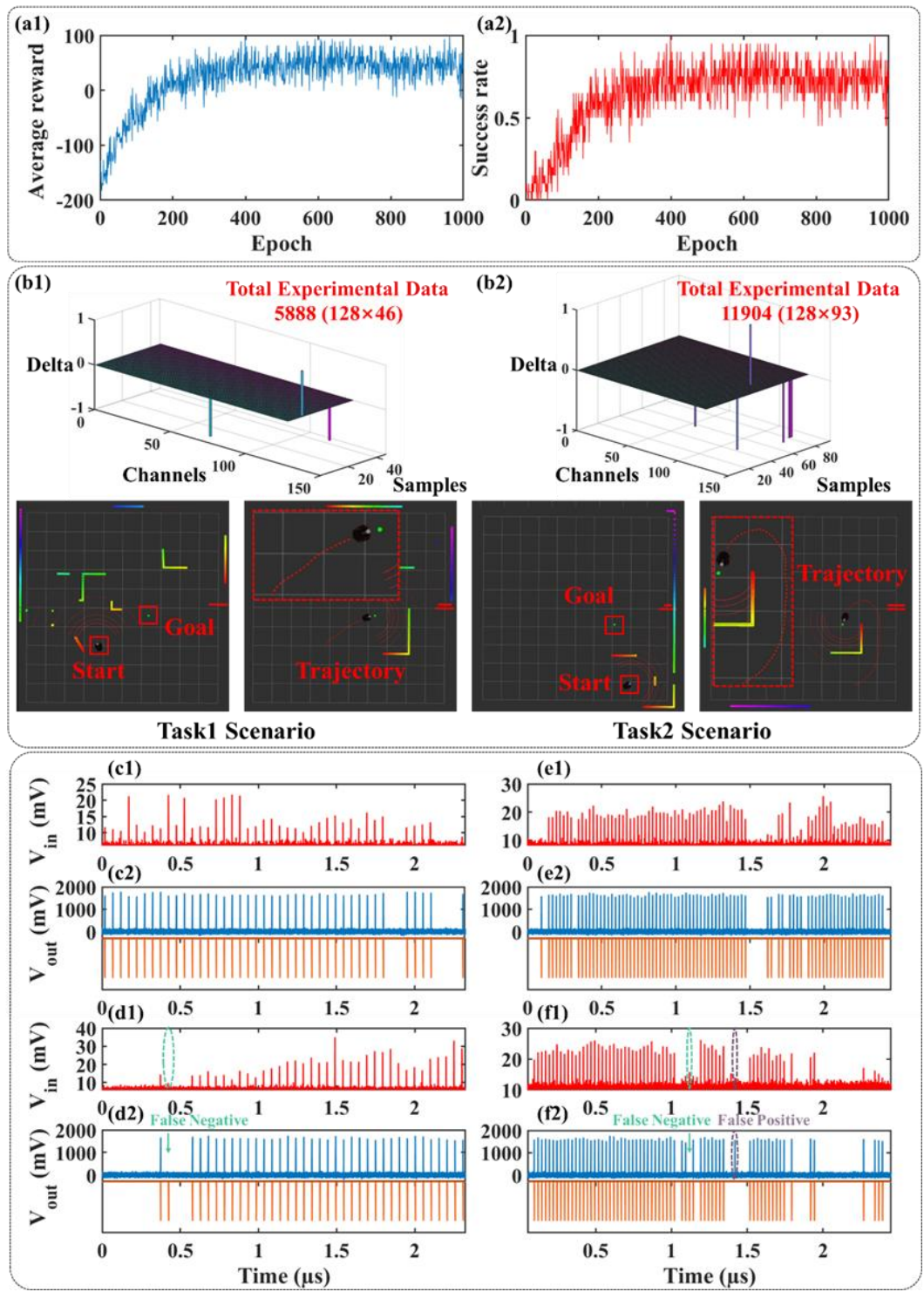

**Figure 3.** Software training results and experimental nonlinear outputs for autonomous navigation tasks. (a1) - (a2): Average reward and success rate for software training. (b1) - (b2): Distributions of discrepancies between the expected spiking activation outputs and the experimentally measured spiking activation outputs for two task scenarios. (c1) - (f1): Comparative examples of experimental inputs and outputs for two task scenarios. The red curves represent the experimental inputs, the blue curve represent the experimentally measured spiking activation outputs, and the orange curve represents the expected spiking activation outputs.

In Task 1, only 3 discrepancies were observed among 5,888 samples (128×46), yielding a hardware-software co-inference error rate of 0.051%. In Task 2, 7 discrepancies occurred among 11,904 samples (128×93), corresponding to an error rate of 0.059%. Figures 3(c1) - (f2) present representative input-output comparisons for the two task scenarios. For Task 1, Figs. 3(c1) - (c2) display a correctly matched spiking activation in channel 1, while Figs. 3(d1) - (d2) show a mismatch in channel 69, where the DFB-SA laser failed to generate a spike at 0.43 μs. For Task 2, Figs. 3(e1) - (e2) illustrate a correctly matched spiking activation in channel 1, whereas Figs. 3(f1) - (f2) reveal an overactivated spike at 1.12 μs and a missing spike at 1.42 μs in channel 114. The proposed architecture demonstrates good scalability and can be extended to a wide range of navigation tasks beyond the two tested scenarios. Complete input-output comparisons for all mismatched activations are provided in SI-4.

Furthermore, we evaluate the inference latency and energy consumption of the proposed photonic spiking TD3 reinforcement learning architecture. The power consumption of a single DFB-SA laser chip is estimated as $1.097 \times 27.9 + 0.76 \times 1.93 = 32.07$ mW, corresponding to a total power consumption of $P_{\text{DFB-SA}} = 4.10$ W for 128 devices, which demonstrates its low-power advantage [52]. The self-pulsation frequency of a single DFB-SA laser chip is 5.23 GHz, corresponding to an electro-optic delay of $\tau_{\text{DFB-SA}} = 1 / 5.23$ GHz $= 191.20$ ps [66]. Accordingly, the photonic spiking TD3 reinforcement learning architecture achieves an inference latency of 191.20 ps per inference, with an energy consumption of $E_{\text{inf}} = P_{\text{DFB-SA}} \times \tau_{\text{DFB-SA}} = 0.78$ nJ/inf. Compared with the 2083 ps/inf inference latency of the Eyeriss architecture[67] and the 11.98 μJ/inf energy consumption of the PopSAN architecture[68], the proposed approach exhibits substantially reduced latency and markedly improved energy efficiency.

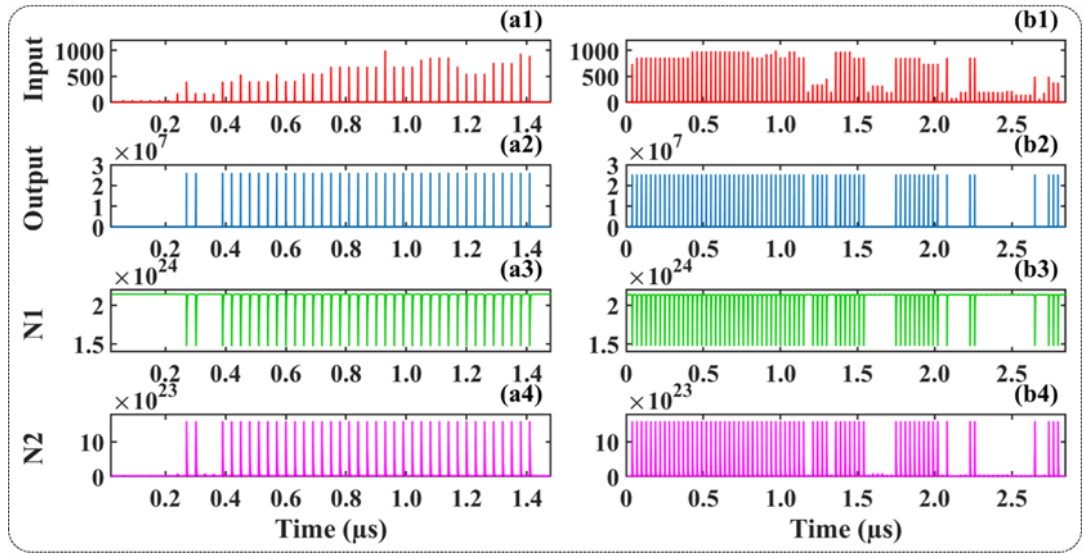

**Figure 4.** Simulation results of the DFB-SA laser corresponding to the input data of error-activated channels for autonomous navigation tasks. (a1) - (a4): simulation results for Channel 69 in Task 1; (b1) - (b4): simulation results for Channel 114 in Task 2. (a1), (b1): externally injected optical signals; (a2), (b2): the outputs of the DFB-SA laser; (a3), (b3): the carrier density in the gain region; (a4), (b4): the carrier density in the saturable absorber.

Finally, based on the experimental results, we further simulate the algorithm input data corresponding to the error-activated channels using the Yamada model to theoretically validate the dynamic characteristics of the DFB-SA laser. Representative simulation results are shown in Fig. 4, where Figs. 4(a1) - (a4) correspond to Channel 69 in Task 1, and Figs. 4(b1) - (b4) correspond to Channel 114 in Task 2. As can be seen, the simulated spiking activations exhibit full agreement with the expected outputs, confirming the theoretical feasibility of achieving accurate DFB-SA laser activation. Complete simulation results for all error-activated channels in both tasks are provided in SI-5.

## 4. Discussion

In this section, we further investigate the performance of PSNNs with larger network scales and longer timesteps. To ensure a fair comparison, only the hidden-layer dimension and timestep length of the photonic spiking Actor network are varied, while all other parameters are kept identical, and the positive-weight constraint applied during photonic hardware-compatible training is removed. For the network-scale analysis, smaller networks showed stronger oscillations and poorer convergence (100×100); therefore, training are conducted with hidden-layer dimensions of 128×128, 256×256, and 512×512 under a fixed timestep of T = 1 for comparison. For the timestep analysis, the hidden-layer dimension was fixed at 128×128, and the timestep was varied as T = 1, 2, and 4. The average reward and navigation success rate curves of the baseline software algorithm are presented in Fig. 5.

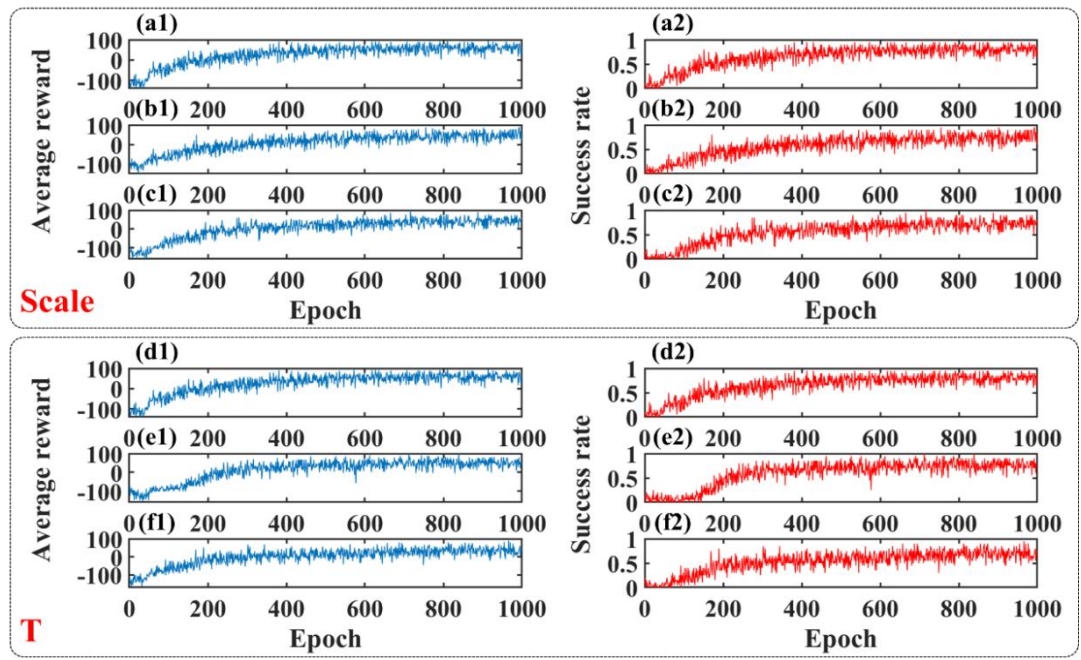

**Figure 5.** Average reward and navigation success rate of software training under different network scales and timesteps. Hidden-layer dimensions: (a1) - (a2), 128×128; (b1) - (b2), 256×256; (c1) - (c2), 512×512. Timesteps: (d1) - (d2), T = 1; (e1) - (e2), T = 2; (f1) - (f2), T = 4.

Subsequently, we analyze the training results. Because oscillations appear in the curves after convergence, the mean and standard deviation are computed over the last 20% of the training data, and the results are summarized in the form of "mean ± standard deviation," as shown in Table 2. The results show that neither increasing the hidden-layer dimension nor extending the timestep effectively improves the performance of the PSNN. On the contrary, both factors tend to destabilize the training process, resulting in more pronounced oscillations in the reward curves. Possible explanations include gradient vanishing or explosion during backpropagation, overfitting caused by the larger number of network parameters, and degraded performance due to hardware resource limitations. Although larger networks and longer timesteps occasionally achieve higher peak performance—reaching up to 100% navigation accuracy in certain epochs—they also yield lower reward values during the early stages of training. In contrast, a hidden-layer dimension of 128×128 with a time step of T = 1 achieves stable and efficient training for autonomous navigation tasks while minimizing hardware resource demands.

**Table 2.** The training results for different hidden-layer dimensions and timesteps

| Ablation study | Parameter | Average reward | Success rate |
|----------------|-----------|----------------|--------------|
| Scale | $100 \times 100$ | $37.73 \pm 20.58$ | $69.7\% \pm 10.1\%$ |
| | $128 \times 128$ | $58.22 \pm 17.29$ | $80\% \pm 8.3\%$ |
| | $256 \times 256$ | $44.60 \pm 20.97$ | $74\% \pm 10.3\%$ |
| | $512 \times 512$ | $40.25 \pm 19.26$ | $71.6\% \pm 9.5\%$ |
| T | 1 | $58.22 \pm 17.29$ | $80\% \pm 8.3\%$ |
| | 2 | $48.82 \pm 18.71$ | $75.8\% \pm 9.1\%$ |
| | 4 | $35.36 \pm 20.92$ | $69.4\% \pm 10\%$ |

## 5. Conclusion

This work proposed and experimentally validated a photonic spiking TD3 reinforcement learning architecture for neuromorphic autonomous navigation, enabling highly energy-efficient intelligent decision-making through hardware-software co-inference. The architecture integrated a spiking-based Actor network with dual continuous-valued Critic networks and successfully completed navigation tasks in the Gazebo simulation environment, achieving an average reward of 58.22±17.29 and a navigation success rate of 80%±8.3%. The final nonlinear spiking activation layer of the photonic Actor network was implemented on a DFB-SA laser array. The proposed system achieved an estimated inference energy of 0.78 nJ/inf and an ultra-low latency of 191.20 ps/inf, highlighting the intrinsic low-power and low-latency advantages of photonic computing for reinforcement learning. Hardware–software collaborative inference further demonstrated high accuracy, with error rates as low as 0.051% and 0.059% under obstacle-free and obstacle-interference scenarios, respectively. In addition, simulations based on the Yamada model were conducted for error-activated channels using algorithm-generated, showing full agreement with the expected spiking responses and thereby theoretically validating the dynamic characteristics of the DFB-SA laser. These results establish a solid foundation for integrating the proposed architecture with large-scale photonic linear computing chips to achieve fully-functional photonic computation. Ablation studies on hidden-layer dimensions and timesteps further confirmed the rationality of the adopted network configuration. While the present work focuses on relatively simple navigation tasks, future efforts will target large-scale system integration, multi-task generalization, and optoelectronic co-packaging. Leveraging the inherent parallelism and ultrafast response of photonic computing, the proposed approach offers a promising pathway toward next-generation real-time neuromorphic autonomous navigation systems.

**Data Availability**

The data that support the findings of this study are available from the corresponding author upon reasonable request.

**Corresponding Author**

*Email: syxiang@xidian.edu.cn

*Email: zhangyahui@xidian.edu.cn

**Supporting Information**

Supporting Information is available from the Wiley Online Library or from the author.

**Acknowledgements**

This work was supported by the National Natural Science Foundation of China

(No.62535015, 62575231); The Fundamental Research Funds for the Central Universities (QTZX23041); Xidian University Specially Funded Project for Interdisciplinary Exploration (TZJH2024009).

**Competing Financial Interests**

The authors declare no competing financial interest.

**Biographies**

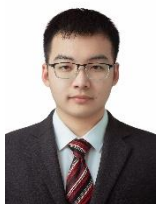

**Yonghang Chen** was born in Shaanxi Province, China, in 2002. He is pursuing a Master's degree at Xidian University in Xi'an, China. His research interests include photonic neuromorphic computing and photonic spiking neural networks.

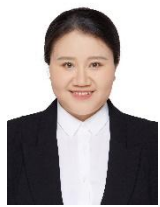

**Shuiying Xiang** was born in Jiangxi Province, China, in 1986. She received the Ph.D. degree from Southwest Jiaotong University, Chengdu, China, in 2013.

She is currently Professor with State Key Laboratory of Integrated Service Networks, Xidian University, Xi'an, China. She is the author or coauthor of more than 140 research papers. Her research interests include vertical cavity surface-emitting lasers, neuromorphic photonic systems, brain-inspired information processing, chaotic optical communication, and semiconductor lasers dynamics.

**Xintao Zeng** was born in Fujian, China, in 1998. He is currently working toward the B.S. degree from Xidian University, Xi'an, China. His research interests include photonic neuromorphic computing, photonic spiking neural networks, optoelectronic chips and integration technologies, and silicon optoelectronic chips and integration technologies.

**Mengting Yu** was born in Jiangxi Province, China, in 2002. She is pursuing a Master's degree at Xidian University in Xi'an, China. Her research interests include silicon photonic MZI chips and photonic spiking reinforcement learning based on robotics.

**Tao Zou** was born in Sichuan Province, China, in 2000. He is pursuing a Master's degree at Xidian University in Xi'an, China. His research interests include FPGA, brain-inspired information processing and spiking neural network.

**Shangxuan Shi** was born in Shanxi Province, China, in 2000 and is currently pursuing the M.S. degree at Xidian University, Xi'an, China, with research interests in photonic neuromorphic computing and photonic spiking neural networks.

**Xingxing Guo** was born in Ji'an City, Jiangxi Province, China, in 1993. She received the Ph.D. degree from Xidian University, Xi'an, China. Her research interests include the vertical cavity surface-emitting lasers, reservoir computing, and chaos communication.

**Yanan Han** was born in Ningxia Hui Autonomous Region, China, in 1996. She received the Ph.D. degree with Xidian University, Xi'an, China. Her research interests include the dynamics and applications of semiconductor lasers, random number generators, and brain-inspired information processing.

**Yahui zhang** was born in Hebei Province, China, in 1993. She received the Ph.D. degree from Xidian University, Xi'an, China. Her research interests include the neuromorphic photonic systems, brain-inspired information processing and random number generators.

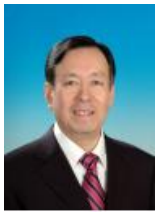

**Yue Hao** was born in the city of Chongqing, China, in 1958. He received the Ph.D. degree from Xi'an Jiaotong University, Xi'an, China, in 1991. He is senior member of the IEEE, executive director of the Chinese Association of Electronics, chairman of the executive councils of the Shaanxi Provincial Association of Electronics, the Trade Association of Integrated Circuits, and the Shaanxi Provincial Semiconductor Illumination Association. He is the leader of the experts group for the implementation of the major sci-tech items of "core electronic devices, high-end universal chips and basic software products" in the medium-to-long term program. He is the leader of the microelectronic technology experts group of the General Armament Department of the People's Liberation Army of China. He is the vice chairman of the national steering committee of the specialty of electronic information science and engineering. Member of the Ninth and Tenth Chinese People's Political Consultative Conference and deputy to the Eleventh National People's Congress of the People's Republic of China. He is Vice president of Xidian University, State Key Discipline Laboratory of Wide Bandgap Semiconductor Technology, School of Microelectronics, Xidian University, Xi'an, China. His current research interests include wide forbidden band semiconductor materials and devices.